\documentstyle[preprint,prl,aps]{revtex}

\def\beq{\begin{equation}}
\def\eeq{\end{equation}}
\def\beqa{\begin{eqnarray}}
\def\eeqa{\end{eqnarray}}
\def\lla{\left\langle}
\def\rra{\right\rangle}
\def\za{\alpha}
\def\zb{\beta}
\def\lsim{\mathrel{\raise.3ex\hbox{$<$\kern-.75em\lower1ex\hbox{$\sim$}}} }
\def\gsim{\mathrel{\raise.3ex\hbox{$>$\kern-.75em\lower1ex\hbox{$\sim$}}} }

\def\smiley{\hbox{\large$\bigcirc$\hspace{-.80em}%
\raise.2ex\hbox{$\cdot\cdot$}\kern-.61em  
\lower.2ex\hbox{\scriptsize$\smile$}}\ }
\def\frowney{\hbox{\large$\bigcirc$\hspace{-.80em}%
\raise.2ex\hbox{$\cdot\cdot$}\kern-.635em
\lower.2ex\hbox{\scriptsize$\frown$}}\ }
\def\blahey{\hbox{\large$\bigcirc$\hspace{-.80em}%
\raise.2ex\hbox{$\cdot\cdot$}\kern-.46em
\lower.3ex\hbox{\scriptsize\hbox{--}}}\ }
\begin{document}


\preprint{{\vbox{\hbox {UR-1523} \hbox{April 1998}}}}



\title{A Simple Phenomenological Parametrization of Supersymmetry 
without R-Parity}


\author{\bf Mike Bisset, Otto C. W. Kong, Cosmin Macesanu, and Lynne H. Orr
\footnote{E-mails: bisset@urhepf.pas.rochester.edu\ ; kong@pas.rochester.edu
\ ;\\ \hspace*{.7in} mcos@pas.rochester.edu \ ; orr@urhep.pas.rochester.edu \ .}
}
\address{Department of Physics and Astronomy,\\
University of Rochester, Rochester NY 14627-0171}
\maketitle

\vskip -0.5cm

\begin{abstract}
We present a parametrization of the supersymmetric standard model without 
R-parity that permits efficient phenomenological analyses of the full model 
without {\it a priori} assumptions. Under the parametrization, which is 
characterized by a single vacuum expectation value for the scalar components 
of the $Y=-1/2$ superfields, the expressions for tree-level mass matrices are 
quite simple. They do not involve the trilinear R-parity violating couplings;
however, the bilinear ${\mu}_i$ terms do enter and cannot be set to zero 
without additional assumptions. We set up a framework for doing phenomenology 
and show some illustrative results for fermion mass matrices and related 
bounds on parameters.  We find in particular that large values of $\tan\!\zb$ 
can suppress R-parity violating effects, substantially weakening 
experimental constraints.  

\end{abstract}

\newpage

\section{Introduction}

A softly broken supersymmetry (SUSY) around or below the scale of
a TeV is no doubt the most popular extension of the Standard Model.
Most SUSY studies concentrate on a $``$minimal" version of such a model 
which contains two  electroweak symmetry breaking Higgs doublets and an 
{\it ad hoc} discrete symmetry, called R-parity, which essentially 
distinguishes particles from superparticles. The phenomenological role of
R-parity is to forbid B- or L-number violating couplings for which
there are important experimental bounds, for instance those from 
superparticle mediated proton decays. However, strict R-parity conservation 
is not required to satisfy these bounds. Furthermore, allowing R-parity
to be broken, either spontaneously through sneutrino VEV's or
explicitly in the Lagrangian, gives rise to interesting phenomenology. 
This has become a subject of much interest recently (for some recent reviews 
and references, see \cite{rpv,Valle}). Because of the large number of possible 
R-parity violating (RPV) couplings, most studies impose assumptions to 
restrict the analyses to a particular subset of RPV couplings.
We wish to adopt a purely phenomenological point of view that allows
an analysis of all the RPV couplings without {\it a priori} assumptions.  
In this letter we present our perspective on the parametrization of the 
general RPV Lagrangian and show some initial results.  Here we concentrate on 
tree level results for mass matrices in the fermion sector, especially 
in the large $\tan\!\beta$ regime; a more detailed study will
appear elsewhere.    

\section{Parametrization of R-parity violation}

The most general renormalizable superpotential for the supersymmetric
standard model without R-parity can be written as
\beq
W = \varepsilon_{ab}\left[ \mu_{\za}  \hat{L}_{\za}^a \hat{H}_u^b
+ h_{ik}^u \hat{Q}_i^a   \hat{H}_{u}^b \hat{U}_k^{\scriptscriptstyle C}
+ \lambda_{i\za k}^{'} \hat{Q}_i^a \hat{L}_{\za}^b 
\hat{D}_k^{\scriptscriptstyle C} + \lambda_{\za \zb k}  \hat{L}_{\za}^a  
 \hat{L}_{\zb}^b \hat{E}_k^{\scriptscriptstyle C}
\right] + \lambda_{i jk}^{''}  \hat{D}_i^{\scriptscriptstyle C}  
\hat{D}_j^{\scriptscriptstyle C}  \hat{U}_k^{\scriptscriptstyle C}\; ,
\eeq
where  $(a,b)$ are $SU(2)$ indices, $(i,j,k)$ are family (flavor) indices,
and $(\za,\zb)$ are (extended) flavor indices from $0$ to $3$ with  
$\hat{L}_{\za}$'s denoting the four doublet superfields with $Y=-1/2$.  
$\lambda$ and $\lambda^{''}$ are antisymmetric in the first two indices as
required by  $SU(2)$ and  $SU(3)$ product rules respectively.
In the limit where $\lambda_{ijk}, \lambda^{'}_{ijk},  \lambda^{''}_{ijk}$
and $\mu_{i}$  all vanish, one recovers the expression for the R-parity 
preserving minimal supersymmetric standard model (MSSM), with $\hat{L}_0$ 
identified as $\hat{H}_d$. In that case, the two Higgses acquire vacuum 
expectation values (VEV's) and the lepton and quark Yukawa couplings are 
given by 
$h^{e}_{ik} \equiv 2\lambda_{i0k} (=-2\lambda_{0ik})$, 
$h^{d}_{ik} \equiv \lambda^{'}_{i0k}$, and 
-$h^{u}_{ik}$, respectively. In the general case, the full expression for 
$W$ together with all admissible soft SUSY breaking terms should be used to 
construct the scalar potential and solve for the vacuum. The  solution is 
then expected to involve VEV's for all five neutral scalars, {\it i.e.} for 
both the Higgses and sneutrinos in the usual terminology. 

It is not necessary, however, to retain all five VEV's in parametrizing
the model, because the freedom associated with the choice of flavor basis
creates redundancy amongst the parameters.  For example, in the MSSM, 
the two $3\times 3$ complex mass matrices -$\frac{{v}_{u}}{\sqrt 2} h^{u}_{ik}$
and $\frac{{v}_{d}}{\sqrt 2} h^{d}_{ik}$ correspond only to ten real 
parameters describing the six mass eigenvalues and the CKM-matrix. And 
although there are models attempting to construct the full high-energy mass
matrices\cite{hsy} in  particular flavor bases, so far as low-energy
phenomenology is concerned the different bases cannot be distinguished.
A fruitful strategy in the MSSM is therefore to choose a flavor basis that 
parametrizes the two Yukawa matrices $h^u$ and $h^d$ with exactly ten 
parameters, namely assuming one mass matrix to be  given by the 
diagonal eigenvalues and the other a multiple of the CKM-matrix and the other 
diagonal eigenvalue matrix. Similarly, in parametrizing the general 
supersymmetric standard model without R-parity, $U(3)$ flavor rotations for 
$\hat{Q}_i, \hat{U}^{\scriptscriptstyle C}_i, \hat{D}^{\scriptscriptstyle C}_i$ 
and $\hat{E}^{\scriptscriptstyle C}_i$ as well as a $U(4)$ rotation for 
$\hat{L}_{\za}$ can be exploited.

The above observation is not new. The popular parametrization exploiting
the flavor rotations has all $\mu_\za$'s except $\mu_{\scriptscriptstyle 0}$ 
as well as two of the  sneutrino VEV's ($\tilde{\nu}_i$'s) set to zero 
(rotated away)~\cite{bh}. (In this context sneutrinos refer to the scalar 
components of the three $\hat{L}_i$ superfields as defined in the basis where 
the $\mu_i$'s are zero.) Note that the lepton Yukawa matrix $h^e_{ik}$ 
{\it cannot} then be taken as diagonalized, since the full $\hat{L}_\za$ basis 
is already fixed. This (single-$\mu$)  parametrization was introduced 
originally in studies of RPV effects on the leptons under the assumption that 
the trilinear RPV couplings are zero\cite{bh}. Extending its
usage to the most general RPV scenario proves difficult. For instance,
the chargino-charged-lepton mass matrix in a generic basis is given by
\beq
{\cal{M}_{\scriptscriptstyle C}} =
 \left(
{\begin{array}{ccc}
{M_{\scriptscriptstyle 2}} &  
\frac{g_{\scriptscriptstyle 2}{v}_{\scriptscriptstyle u}}{\sqrt 2}  & 0  \\
 \frac{g_{\scriptscriptstyle 2}{v}_{\scriptscriptstyle d}}{\sqrt 2} &
 {{ \mu}_{\scriptscriptstyle 0}} &   
 -h^e_{ik}  \frac{\tilde{\nu}_{i}}{\sqrt 2} \\
\frac{g_{\scriptscriptstyle 2}{v}_{\scriptscriptstyle i}}{\sqrt 2} 
& {{ \mu}_{i}} &
 h^e_{ik}  \frac{{v}_{\scriptscriptstyle d}}{\sqrt 2} 
+ 2\lambda_{ijk}\frac{\tilde{\nu}_{i}}{\sqrt 2}
\end{array}}
 \right) \; .
\eeq
It is easy to see that under the single-$\mu$ parametrization a set
of $\lambda$-couplings associated with the nonzero sneutrino VEV still
remain, making the analysis of the mass eigenstates quite complicated.
Analogously, $\lambda^{'}$-couplings will enter the down-quark mass
matrix along with the nonzero sneutrino VEV. It has also been pointed
out that rotating away the $\mu_i$'s does not simplify the analysis
of the scalar potential because the RPV soft mass terms also contribute
to the same quadratic terms\cite{mu}.
  
We wish to find a relatively simple parametrization not requiring 
{\it a priori} assumptions. We propose here to use the $U(4)$ rotation to set 
all sneutrino VEV's ($\frac{\tilde{\nu}_{i}}{\sqrt{2}} \, \equiv \, \lla 
\hat{L}_i \rra$) to zero, leaving a single VEV for $\lla \hat{L}_0 \rra$.  
We keep all of the $\mu_{\za}$ while the rest of the leptonic flavor rotations 
are used to set  $h^e_{ik}$ diagonal. In our new basis, the (tree-level)
mass matrices for {\it all} the fermions {\it do not} involve any trilinear 
RPV couplings\footnote{The scalar mass matrices are also much simplified as 
there are only two non-zero VEV's, $v_{\scriptscriptstyle d}$ and 
$v_{\scriptscriptstyle u}$.  Further details will be discussed in \cite{II}.}.
In particular, $\cal{M}_{\scriptscriptstyle C}$ is given by
\beq \label{mc}
{\cal{M}_{\scriptscriptstyle C}} =
 \left(
{\begin{array}{ccccc}
{M_{\scriptscriptstyle 2}} &  
\frac{g_{\scriptscriptstyle 2}{v}_{\scriptscriptstyle u}}{\sqrt 2}  
& 0 & 0 & 0 \\
 \frac{g_{\scriptscriptstyle 2}{v}_{\scriptscriptstyle d}}{\sqrt 2} & 
 {{ \mu}_{\scriptscriptstyle 0}} & 0 & 0 & 0 \\
0 &  {{ \mu}_{\scriptscriptstyle 1}} & {{m}_{\scriptscriptstyle 1}} & 0 & 0 \\
0 & {{ \mu}_{\scriptscriptstyle 2}} & 0 & {{m}_{\scriptscriptstyle 2}} & 0 \\
0 & {{ \mu}_{\scriptscriptstyle 3}} & 0 & 0 & {{m}_{\scriptscriptstyle 3}}
\end{array}}
\right) \; ,
\eeq 
where $m_i = h^e_{ii} \frac{{v}_{\scriptscriptstyle d}}{\sqrt 2}$. The quark 
mass matrix for each sector involves only one VEV and therefore assumes the 
same form as in the MSSM. The neutralino-neutrino mass matrix is then given by
\beq \label{mn}
{\cal{M}_{\scriptscriptstyle N}} 	
=  \left(
{\begin{array}{ccccccc}
{{M}_{\scriptscriptstyle 1}} & 0 &  \frac {{g}_{1}{v}_{u}}{2}
 &  -\frac{{g}_{1}{v}_{d}}{2} & 0 & 0 & 0 \\
0 & {{M}_{\scriptscriptstyle 2}} &  -\frac{{g}_{2}{v}_u}{2} & 
\frac{{g}_{2}{v}_{d}}{2} & 0 & 0 & 0 \\
 \frac {{g}_{1}{v}_{u}}{2} &   -\frac{{g}_{2}{v}_u}{2} & 0 & 
 - {{\mu}_{\scriptscriptstyle 0}} &  - {{ \mu}_{\scriptscriptstyle 1}} 
 &  - {{ \mu}_{\scriptscriptstyle 2}} &  - {{ \mu}_{\scriptscriptstyle 3}} \\
 -\frac{{g}_{1}{v}_{d}}{2} & \frac{g_{2}{v}_d}{2}
 &  - {{ \mu}_{0}} & 0 & 0 & 0 & 0 \\ 
0 & 0 &  - {{ \mu}_{\scriptscriptstyle 1}} & 0 & 0 & 0 & 0 \\
0 & 0 &  - {{ \mu}_{\scriptscriptstyle 2}} & 0 & 0 & 0 & 0 \\
0 & 0 &  - {{ \mu}_{\scriptscriptstyle 3}} & 0 & 0 & 0 & 0
\end{array}}
 \right)  \; ,
\eeq
The simplicity of the  mass matrix expressions is obvious. We would like to 
re-emphasize that this is achieved {\it without any a priori assumptions}; we 
have simply chosen to parametrize the model in a specific flavor 
basis~\cite{Nar}. In our basis, we identify the Higgs $\hat{H}_1 
(\equiv \hat{L}_0)$ as the $Y=-1/2$ doublet that bears the full
VEV among the $\hat{L}_{\za}$'s, and in the interesting region of relatively
small $\mu_i$'s our three $L_i$'s align well with the charged-lepton mass
eigenstates. 

Our single-VEV parametrization helps to simplify analysis of the
model in both the fermion and scalar sectors. We will take advantage of
this to illustrate below some novel features of the model in the large 
$\tan\!{\beta}$ regime. The parametrization also provides a framework that  
easily allows a full phenomenological analysis of the model, with both 
bilinear and trilinear RPV-terms admitted. Finally, all parameters are assumed 
to be real --- potentially rich CP violating features of the model
are not considered here. 

Before going on to our analysis, it is worthwhile to further clarify
some issues about the parametrization of R-parity violation.
The single-VEV parametrization, with explicit bilinear
RPV-terms, does not manifestly exhibit sneutrino VEV's.  Nevertheless,
it is possible to use this framework to describe a spontaneously
broken R-parity scenario, which is defined by the existence of
sneutrino VEV's in some basis where {\it all} explicit RPV terms vanish.
A rotation of the $\hat{L}_{\za}$ would connect such a basis to ours.
Note that if we perform such a rotation to our single-VEV basis, in 
general not only bilinear (${\mu}_i$) terms but also trilinear and 
soft SUSY breaking RPV-terms will be introduced as well. A similar perspective 
holds for models with R-parity broken spontaneously via VEV(s) of extra 
singlet superfield(s) at some higher scale. In that case, the extra 
superfield(s) have to be integrated out to recover the supersymmetric standard 
model. Such models are naturally formulated in a mixed 
parametrization~\cite{Valle}, {\it i.e.} with both $\mu_i$'s and 
$\lla \hat{L}_i \rra$'s.

There are also discussions in the literature which consider the MSSM with
only a few RPV-terms added while the $``$sneutrino VEV's" are assumed
to be zero. One should be particularly careful in interpreting the meaning
of the assumptions in such cases. As implied in the above discussion,
imposing such assumptions at the Lagrangian level is not a (flavor) basis or
parametrization independent procedure. Combining such assumptions with a 
specific choice of basis is a potential source of confusion and sometimes  
inconsistency. In our opinion, a clear interpretation is
provided by beginning with the Lagrangian in the single-VEV parametrization
we proposed above.  A given model is then identified by specifying  which 
RPV-terms are not admitted. An important related point is that
so long as the $\mu_i$'s are not all assumed to be zero, the $L_i$'s are
not  exactly indentifiable as the physical charged leptons, nor are the 
charginos and neutralinos, for instance, the same states as in the R-parity 
conserving MSSM. The vanishing of the $\mu_i$'s, if taken,  would be an 
assumption. All in all, while specific RPV models may be more naturally 
formulated in a particular parametrization, the single-VEV parametrization, we 
believe, provides a particularly efficient framework for a model 
independent study of constraints on and phenomenology of R-parity violation.

\section{Analysis of neutral fermion mass matrix}

The generation of non-zero neutrino mass(es) is one of the most prominent 
features of R-parity violation. As a result the experimental neutrino
mass bound has been used to put constraints on various RPV-couplings~\cite{nu}. 
Two neutrino eigenstates are left massless at the tree level, while the 
third one gains a mass through the RPV-couplings ($\mu_i$'s) to the higgsino, 
as can be  seen from Eq.(\ref{mn}). Note that the massive eigenstate is in 
general a mixture of all three neutrino states. In fact one  can use a simple 
rotation to decouple the massless states. The remaining $5\times 5$ mass 
matrix is then given by
\beq \label{mn5}
{\cal{M}_{\scriptscriptstyle N}}^{\!\!\scriptscriptstyle (5)} 	
=  \left(
{\begin{array}{ccccc}
{M}_{\scriptscriptstyle 1} & 0 &  \frac {{g}_{1}{v}_{u}}{2}
 &  -\frac{{g}_{1}{v}_{d}}{2} & 0 \\
0 & {{M}_2} &  -\frac{{g}_{2}{v}_u}{2} 
& \frac{{g}_{2}{v}_{d}}{2} & 0 \\
 \frac {{g}_{1}{v}_{u}}{2} &   -\frac{{g}_{2}{v}_u}{2} & 0 &  
 - {{\mu}_{\scriptscriptstyle 0}} &  - {{\mu}_{\scriptscriptstyle 5}}  \\
 -\frac{{g}_{1}{v}_{d}}{2} & \frac{g_{2}{v}_d}{2}
 &  - {{\mu}_{\scriptscriptstyle 0}} & 0 & 0  \\ 
0 & 0 &  - {{\mu}_{\scriptscriptstyle 5}} & 0 & 0 
\end{array}} \right) \; ,
\eeq
where 
\beq
\mu_{\scriptscriptstyle 5} = \sqrt{\mu_{\scriptscriptstyle 1}^2 
+\mu_{\scriptscriptstyle 2}^2 +\mu_{\scriptscriptstyle 3}^2} \; ;
\eeq
and the corresponding massive neutrino state is given by
\beq
\left|\nu_{\scriptscriptstyle 5}\rra = \frac{\mu_{\scriptscriptstyle 1}}
{\mu_{\scriptscriptstyle 5}}\left|\nu_{\scriptscriptstyle 1}\rra
+ \frac{\mu_{\scriptscriptstyle 2}}{\mu_{\scriptscriptstyle 5}}
\left|\nu_{\scriptscriptstyle 2}\rra +  \frac{\mu_{\scriptscriptstyle 3}}
{\mu_{\scriptscriptstyle 5}}\left|\nu_{\scriptscriptstyle 3}\rra \; .
\eeq
The common strategy to obtain the neutrino mass corresponds to 
assuming a small $\mu_{\scriptscriptstyle 5}$ where 
${\cal{M}_{\scriptscriptstyle N}}^{\!\!\scriptscriptstyle (5)}$
adopts a $``$seesaw" structure. This gives
\beq
m_{\nu_{\scriptscriptstyle 5}} = 
\frac{\det{\cal{M}_{\scriptscriptstyle N}}^{\!\!\scriptscriptstyle (5)}}
{\det{{\cal{M}}_{\scriptscriptstyle 4\times 4}}} =  -  \frac {1}{2}
\frac{ {\mu}_{5}^{2} {v}^{2} \cos^2\!\!\zb 
\left( x{g}_{\scriptscriptstyle 2}^{2} 
+ {g}_{\scriptscriptstyle 1}^{2} \right) }
{\mu_{\scriptscriptstyle 0} \left[ 2xM_{\scriptscriptstyle 2}
 \mu_{\scriptscriptstyle 0} -
 \left( x{g}_{\scriptscriptstyle 2}^{2}+{g}_{1\scriptscriptstyle }^{2}\right) 
{v}^2 \sin\!{\zb}\cos\!{\zb} \right] }  
\eeq
where we have substituted $v_{\scriptscriptstyle d}=v\cos\!{\zb}$, 
$v_{\scriptscriptstyle u}=v\sin\!{\zb}$,
and $M_{\scriptscriptstyle 1}=xM_{\scriptscriptstyle 2}$. Note that for large 
$\tan\!{\zb}$, $\cos\!{\zb}$ is a suppression factor. For example, at 
$\tan\!{\zb}=45$, saturation of the machine bound of 
$24\, \hbox{MeV}$~\cite{PDG} for $m_{\nu_{\scriptscriptstyle 5}}$ allows a 
$\mu_{\scriptscriptstyle 5}$ value as large as the chargino mass scale 
$M_{\scriptscriptstyle 2}$ and the higgsino mass mixing parameter, 
$\mu_{\scriptscriptstyle 0}$. Now, large $\mu_{\scriptscriptstyle 5}$ values 
are beyond the validity of the  $``$seesaw" analysis; however, an alternative 
perturbative analysis can be performed treating the EW-symmetry breaking 
terms in ${\cal{M}_{\scriptscriptstyle N}}^{\!\!\scriptscriptstyle (5)}$ as a
perturbation.  The first order part can then be diagonalized exactly 
without any assumptions about the  magnitude of $\mu_{\scriptscriptstyle 5}$. 
The resulting zero eigenvalue is lifted by the perturbation to give
\beq \label{mnu}
m_{\nu_{\scriptscriptstyle 5}}= - \frac{1}{4} 
\frac{{\mu}_{\scriptscriptstyle 5}^{2} {v}^{2} \cos\!\!^2\zb 
\left(x{g}_{\scriptscriptstyle 2}^{2} + {g}_{\scriptscriptstyle 1}^{2} \right)}
{ \left( \mu_{\scriptscriptstyle 0}^{2} + \mu_{\scriptscriptstyle 5}^{2} 
\right) xM_{\scriptscriptstyle 2} }\; ,
\eeq
with the eigenvector in the original basis given by
\beq
 \left(\begin{array}{ccccc}
 \frac{{\mu}_{5} {g}_{1} v\cos\!{\zb}}{2 xM_{\scriptscriptstyle 2}} 
&  -  \frac{{\mu}_{5} {g}_{2} v\cos\!{\zb}}{2 M_{\scriptscriptstyle 2}} 
 & 0 &  -  \mu_{\scriptscriptstyle 5}  &   \mu_{\scriptscriptstyle 0}
\end{array} \right) \; .
\eeq
One can rewrite  Eq.(\ref{mnu}) to obtain a bound on 
$\mu_{\scriptscriptstyle 5}$:
\beq \label{mu5}
\mu_{\scriptscriptstyle 5}^2
< \frac {4 {x}{\mu}_{\scriptscriptstyle 0}^{2}M_{\scriptscriptstyle 2} 
m_{\nu_{\scriptscriptstyle 5({\text b\!o\!u\!n\!d})}}}{{v}^{2} \cos^{2}\!\!{\zb}
\left( x{g}_{\scriptscriptstyle 2}^{2} + {g}_{\scriptscriptstyle 1}^{2} \right) 
- 4{x}M_{\scriptscriptstyle 2}  
m_{\nu_{\scriptscriptstyle 5({\text b\!o\!u\!n\!d})}}} \; .
\eeq
As $M_{\scriptscriptstyle 2}$ increases, the denominator above drops to zero,
beyond which there is {\it no}  bound on $\mu_{\scriptscriptstyle 5}$. 
For $\tan\!{\zb}=45$, this happens at 
$M_{\scriptscriptstyle 2}  \sim  210\, \hbox{GeV}$.
We note that in order to
use the $\nu_\tau$ machine mass bound of $24\, \hbox{MeV}$
for $m_{\nu_{\scriptscriptstyle 5}({\text bound})}$,
we have assumed ${\mu}_{\scriptscriptstyle 1}={\mu}_{\scriptscriptstyle 2}= 0$, 
{\it i.e.} ${\mu}_{\scriptscriptstyle 5} = {\mu}_{\scriptscriptstyle 3}$ 
(${\mu}_{\scriptscriptstyle 1}:{\mu}_{\scriptscriptstyle 2}:{\mu}_{\scriptscriptstyle 3} = 0:0:1$).

The perturbative result in Eq.(\ref{mu5}) is borne out by exact numerical 
results from diagonalizing the neutral fermion mass matrix, as illustrated in 
Figure 1, for $\tan\!{\zb}=45$; the corresponding result for $\tan\!{\zb}=2$ 
is also shown for comparison. The difference between 
${\mu}_{\scriptscriptstyle 5}$ bounds for the two cases is striking.  In both 
cases we see that the neutrino mass bound on ${\mu}_{\scriptscriptstyle 5}$ 
($={\mu}_{\scriptscriptstyle 3}$ here) is tighter for low values of 
$|{\mu}_{\scriptscriptstyle 0}|$ and weakens as $|{\mu}_{\scriptscriptstyle 0}|$
is increased. For $\tan\!{\zb}=45$, values of ${\mu}_{\scriptscriptstyle 5}$ 
in the hundreds of GeV are completely consistent with the  $24\, \hbox{MeV}$ 
bound for viable regions of the 
$(M_{\scriptscriptstyle 2},{\mu}_{\scriptscriptstyle 0})$ parameter space.
And again, for $M_{\scriptscriptstyle 2}$ large enough, we get no limit on  
${\mu}_{\scriptscriptstyle 5}$ at all.

There are potentially much stronger bounds on  neutrino masses from
cosmological considerations which however depend on the decay modes
and other assumptions so that a neutrino mass above an MeV 
is not definitely ruled out~\cite{cos}. There are also other experimental
constraints  on neutrino masses and mixings. In this first paper, our purpose 
is simply to illustrate the advantage of performing the analysis in the 
single-VEV parametrization as well as to point out the interesting suppression 
of RPV effects in the large $\tan\!\zb$ regime. 

\section{Constraints from chargino-charged lepton mass matrix}

How are the $\mu_{i}$'s otherwise  constrained particularly in the
large $\tan\!{\zb}$ regime? Couplings of the charged leptons to the $Z^0$ are 
well measured and can constrain the $\mu_{i}$'s through the chargino-charged 
lepton mass matrix. In Ref.\cite{NP}, various $Z^0$-couplings constraints on  
R-parity violation are studied assuming the trilinear RPV couplings vanish.
We follow basically the same strategy (but without the latter assumption)
and our  parametrization allows the constraints to be cast
explicitly in terms of the magnitudes of the $\mu_{i}$.  For the large 
$\tan\!{\zb}$ regime, we find  a weakening of the constraints
as a result of $\cos\!{\zb}$ suppression factor(s) even stronger than 
that illustrated above for the neutrino case. 

We begin with a perturbative approach as in the neutral fermion sector.
After we  diagonalize its upper $2\times 2$ (chargino) block, 
${\cal{M}_{\scriptscriptstyle C}}$ (Eq.(\ref{mc})) consists of
two diagonal blocks of fully diagonal sub-matrices and a lower
$3\times 2$ off-diagonal block containing the $\mu_i$ parameters.
The latter is taken as a perturbation to the diagonal matrix. 
It is then a simple exercise to obtain the matrix elements~\cite{bdi} :
\beqa
U^{\dag}_{\scriptscriptstyle L}(i+2,1) &=& 
- \frac{{{ \mu}_{i}}\sqrt{2}{M_{\scriptscriptstyle W}}
\cos\!{\zb}}{\mu_{\scriptscriptstyle 0} M_{\scriptscriptstyle 2} 
- 2 M_{\scriptscriptstyle W}^2\sin\!{\zb}\cos\!{\zb}} \; ,  \nonumber \\
U^{\dag}_{\scriptscriptstyle L}(i+2,2) &=& 
-\frac{\mu_i M_{\scriptscriptstyle 2}}
{\mu_{\scriptscriptstyle 0} M_{\scriptscriptstyle 2} 
- 2 M_{\scriptscriptstyle W}^2\sin\!{\zb}\cos\!{\zb}} \; , \\
U^{\dag}_{\scriptscriptstyle R}(i+2,1) &=& 
-\frac{m_i{\mu}_{i}\sqrt{2}{M_{\scriptscriptstyle W}}
\left(M_{\scriptscriptstyle 2}\sin\!{\zb}
+\mu_{\scriptscriptstyle 0}\cos\!{\zb}\right)}
{(\mu_{\scriptscriptstyle 0} M_{\scriptscriptstyle 2} 
- 2 M_{\scriptscriptstyle W}^2\sin\!{\zb}\cos\!{\zb})^2} \; ,  \nonumber \\
U^{\dag}_{\scriptscriptstyle R}(i+2,2) &=& - \frac {{m_i { \mu}_{i}}
\left(M_{\scriptscriptstyle 2}^2
+2M_{\scriptscriptstyle W}^2\cos^2\!\!{\zb}\right)}
{(\mu_{\scriptscriptstyle 0} M_{\scriptscriptstyle 2} 
- 2 M_{\scriptscriptstyle W}^2\sin\!{\zb}\cos\!{\zb})^2} \; ,
\eeqa
with
\beq
U^{\dag}_{\scriptscriptstyle L} {\cal{M}_{\scriptscriptstyle C}} 
U_{\scriptscriptstyle R} = {\rm diag}\{ \bar{M}_{c{\scriptscriptstyle 1}},
 \bar{M}_{c{\scriptscriptstyle 2}}, \bar{m}_{\scriptscriptstyle 1}, 
 \bar{m}_{\scriptscriptstyle 2},  \bar{m}_{\scriptscriptstyle 3} \} \; ,
\eeq
where index $i$ refers to one of the three leptonic states. These matrix 
elements are the ones needed for studying leptonic physics. They characterize 
the gaugino and higgsino contents of each leptonic mass eigenstate.

The $Z^0$-boson coupling to the mass eigenstates is given by
\begin{equation}\label{Zll}
{\cal L}^{\scriptscriptstyle Z\bar{\chi}^-\!\chi^-}_{\scriptscriptstyle int} 
\equiv \frac{g_{\scriptscriptstyle 2}}{2\cos\!\theta_w} Z^\mu
\bar{\chi}^-_i \gamma_\mu \left( \tilde{A}^{\scriptscriptstyle L}_{ij}
\frac{1-\gamma_{\scriptscriptstyle 5}}{2}
+ \tilde{A}^{\scriptscriptstyle R}_{ij}\frac{1+\gamma_{\scriptscriptstyle 5}}{2} \right) \chi^-_j \; .
\end{equation}
Using the results above, together with unitarity of the 
$U_{\scriptscriptstyle L}^{\dag}$ and $U_{\scriptscriptstyle R}^{\dag}$
matrices, we have, for large $\tan\!{\zb}$ where formul\ae\ simplify,
\begin{eqnarray}
\tilde{A}^{\scriptscriptstyle L}_{ij} 
&=& \frac{2\mu_i \mu_j M_{\scriptscriptstyle W}^2 \cos^2\!\!\zb}
{\mu_{\scriptscriptstyle 0}^2 M_{\scriptscriptstyle 2}^2} + 
\delta_{ij}(1 - 2\sin\!^2\!\theta_w) \; , \nonumber \\
 \tilde{A}^{\scriptscriptstyle R}_{ij} &=&  \frac{\mu_i \mu_j m_i m_j
 (M_{\scriptscriptstyle 2}^2+ 4M_{\scriptscriptstyle W}^2)}
 {\mu_{\scriptscriptstyle 0}^4 M_{\scriptscriptstyle 2}^2}
- \delta_{ij}2\sin\!^2\!\theta_w \; .
\end{eqnarray}
Note that while Eq.(\ref{Zll}) is  valid for all five charged fermion states, 
the above given formul\ae\ for the couplings 
$\tilde{A}^{\scriptscriptstyle L,R}_{ij}$'s are only for the three 
leptonic states, to which we limit the present discussion. In accordance with  
the perturbational approach, the results contain $``$small" ratios such as 
$\frac{\mu_i}{\mu_{\scriptscriptstyle 0}}$. The deviations from standard 
universality in $\tilde{A}^{\scriptscriptstyle R}_{ij}$ are typically
ignored,  being suppressed by two factors of the 
$\frac{m_i}{\mu_{\scriptscriptstyle 0}}$ mass ratios. But the corresponding 
deviations in $\tilde{A}^{\scriptscriptstyle L}_{ij}$ have a $\cos^2\!\!\zb$ 
suppression, which for $\tan\!\zb =45$, for example, gives a factor of 
$10^{-3}$. The important point to note here, in the large $\tan\!\zb$ regime, 
is that $\tilde{A}^{\scriptscriptstyle R}_{33}$ or  
$\tilde{A}^{\scriptscriptstyle R}_{23}$ could well be significant in comparison
with the $\tilde{A}^{\scriptscriptstyle L}_{ij}$'s. 

With the $\tilde{A}^{\scriptscriptstyle L,R}_{ij}$ formul\ae, it is then 
straightforward to check the constraints different processes involving 
universality violation or FCNC~\cite{NP} impose on the  
$\frac{\mu_i}{\mu_{\scriptscriptstyle 0}}$ ratios. Specifically, we check
coupling universality and left-right asymmetry, as well as tree level
$Z^0\ell_i\ell_k$ couplings through branching ratios of various
$Z^0\longrightarrow 2\ell$, $\mu\longrightarrow 3\ell$ and 
$\tau\longrightarrow 3\ell$ processes. When a constraint allows
the ratios to be larger than unity, it essentially goes away, from the
present perturbative perspective. This happens for all of these constraints 
for a sufficiently large $M_{\scriptscriptstyle 2}$ except the one from
$\mu^- \longrightarrow e^-e^+e^-$, which has a much stronger experimental 
bound. For $\tan\!\zb = 45$,
$M_{\scriptscriptstyle 2} \; \gsim \; 15$-$35\, \hbox{GeV}$ eliminates all but 
the latter process from which we obtain
\beq \label{mu3e}
\frac{| \mu_{\scriptscriptstyle 1} \mu_{\scriptscriptstyle 2} |}
{\mu_{\scriptscriptstyle 0}^2} \lsim 4.7\times 10^{-7} 
M_{\scriptscriptstyle 2}^2 \; .
\eeq 
With an admissible gaugino mass $M_{\scriptscriptstyle 2}$ of the order 
$100\, \hbox{GeV}$, this only surviving constraint gives a numerical bound on 
$\frac{| \mu_{\scriptscriptstyle 1} \mu_{\scriptscriptstyle 2} |}
{\mu_{\scriptscriptstyle 0}^2}\sim 10^{-3}$ only (the signs of the ${\mu}_i$ 
do not affect any of the results presented here). Furthermore, the same 
behavior appears when the $Z^0 \longrightarrow \nu\nu$ width constraint is 
considered.

In the perturbative calculation, the $m_i$'s are approximated by the 
$\bar{m}_i$'s, the physical charged lepton masses. In the exact computations,
we numerically integrate from $\mu_i = 0$ (for which the
$m_i$ are exactly the $\bar{m}_i$) to the final $\mu_i$ values.
This is necessary to find an acceptable set of $m_i$'s that yield the 
correct physical charged lepton masses for a given set of $\mu_i$'s.
We also then find the chargino masses, which now depend on the
$\mu_i$'s.  For example, the minimum $\mu_i$ values required to 
give both chargino masses above $90\, \hbox{GeV}$ for $\tan\!\zb =45$
and ${\mu}_{\scriptscriptstyle 1}:{\mu}_{\scriptscriptstyle 2}:
{\mu}_{\scriptscriptstyle 3} = 0:1:1$ are shown in Figure 2.

The numerical results also bear out the perturbative analysis
of the couplings (in lepton and $Z^0$ decays) mentioned above.  
Bounds from B.R.$(\mu^- \longrightarrow e^-e^+e^-) < 1.0^{-12}$~\cite{PDG} 
are shown in Figure 3.  Here we plot contours of ${\mu}_5$ assuming 
a ratio ${\mu}_{\scriptscriptstyle 1}:{\mu}_{\scriptscriptstyle 2}:
{\mu}_{\scriptscriptstyle 3} = 1:1:0$.  As with the neutrino mass bound,
the limit on ${\mu}_{\scriptscriptstyle 5}$ is more strict for low values of 
$|{\mu}_{\scriptscriptstyle 0}|$. The substantial weakening of the constraint 
for large $\tan\!\zb$ is well illustrated in the figure. Numerical studies of 
the other processes mentioned above give no restriction on the maximum allowed 
values of ${\mu}_{\scriptscriptstyle 5}$ for
$|{\mu}_{\scriptscriptstyle 0}|,M_{\scriptscriptstyle 2} \gsim 10\, \hbox{GeV}$
(again for $\tan\!\zb =45$), as indicated by  the perturbational results. 

\section{Summary}

In summary, we have presented and illustrated with examples the merits of 
the single-VEV parametrization of supersymmetry without R-parity (see
also~\cite{Nar}), which can be used in phenomenological studies without 
requiring specific model-dependent assumptions. Our
analysis as outlined  above also indicates a strong suppression of RPV 
effects from the bilinear $\mu_i$ terms in the large $\tan\!\zb$ regime. 
We note that the trilinear RPV couplings play no role in the analysis 
discussed here, though they are expected to have an important role in the 
other aspects of the model. This is, however, exactly what makes a 
comprehensive analysis of the full model feasible under the parametrization.  
Further details of such an analysis will be reported in future work.

\acknowledgements
\vskip -0.2cm
We thank K.S. Babu, E. Nardi and S. Pakvasa for helpful discussions.  
This work was supported in part by the U.S. Department of Energy,
under grant DE-FG02-91ER40685 and by the U.S. National Science Foundation,
under grant PHY-9600155.

\eject

{\centerline{{\bf Figure Captions}}

\vskip 3.0cm

{\bf Figure 1}:

Maximum allowed values of ${\mu}_{\scriptscriptstyle 5}$ (in GeV) consistent 
with $m_{{\nu}_{\tau}} < 24\, \hbox{MeV}$
(${\mu}_{\scriptscriptstyle 1}:{\mu}_{\scriptscriptstyle 2}:
{\mu}_{\scriptscriptstyle 3} = 0:0:1$).
$M_{\scriptscriptstyle 1} = xM_{\scriptscriptstyle 2}$, with 
$x = \frac{5}{3}{\tan}^2{\theta}_w$ assumed from gaugino unification 
\newline
($M_{\scriptscriptstyle Z} = 91.19\, \hbox{GeV}, \sin\!^2{\!\theta}_w = 0.23$).
The region above or outside of a given contour 
is excluded for ${\mu}_{\scriptscriptstyle 5}$'s above the indicated value.

\vskip 1.5cm

{\bf Figure 2}:

Minimum values of ${\mu}_{\scriptscriptstyle 5}$ (in GeV) required to give
both chargino masses above $90\, \hbox{GeV}$
\newline
(${\mu}_{\scriptscriptstyle 1}:{\mu}_{\scriptscriptstyle 2}
:{\mu}_{\scriptscriptstyle 3} = 0:1:1$).
The area above or outside of a given contour  
has both chargino masses $> \; 90\, \hbox{GeV}$ for 
${\mu}_{\scriptscriptstyle 5}$'s above the indicated value.

\vskip 1.5cm

{\bf Figure 3}:

Maximum allowed values of ${\mu}_{\scriptscriptstyle 5}$ (in GeV) consistent 
with B.R.$(\mu^- \longrightarrow e^-e^+e^-) < 1.0^{-12}$ 
\newline
(${\mu}_{\scriptscriptstyle 1}:{\mu}_{\scriptscriptstyle 2}
:{\mu}_{\scriptscriptstyle 3} = 1:1:0$).
The region above or outside of a given contour 
is excluded for ${\mu}_{\scriptscriptstyle 5}$'s above the indicated value.

\eject

\begin{figure}

\vspace{1in}
\includegraphics{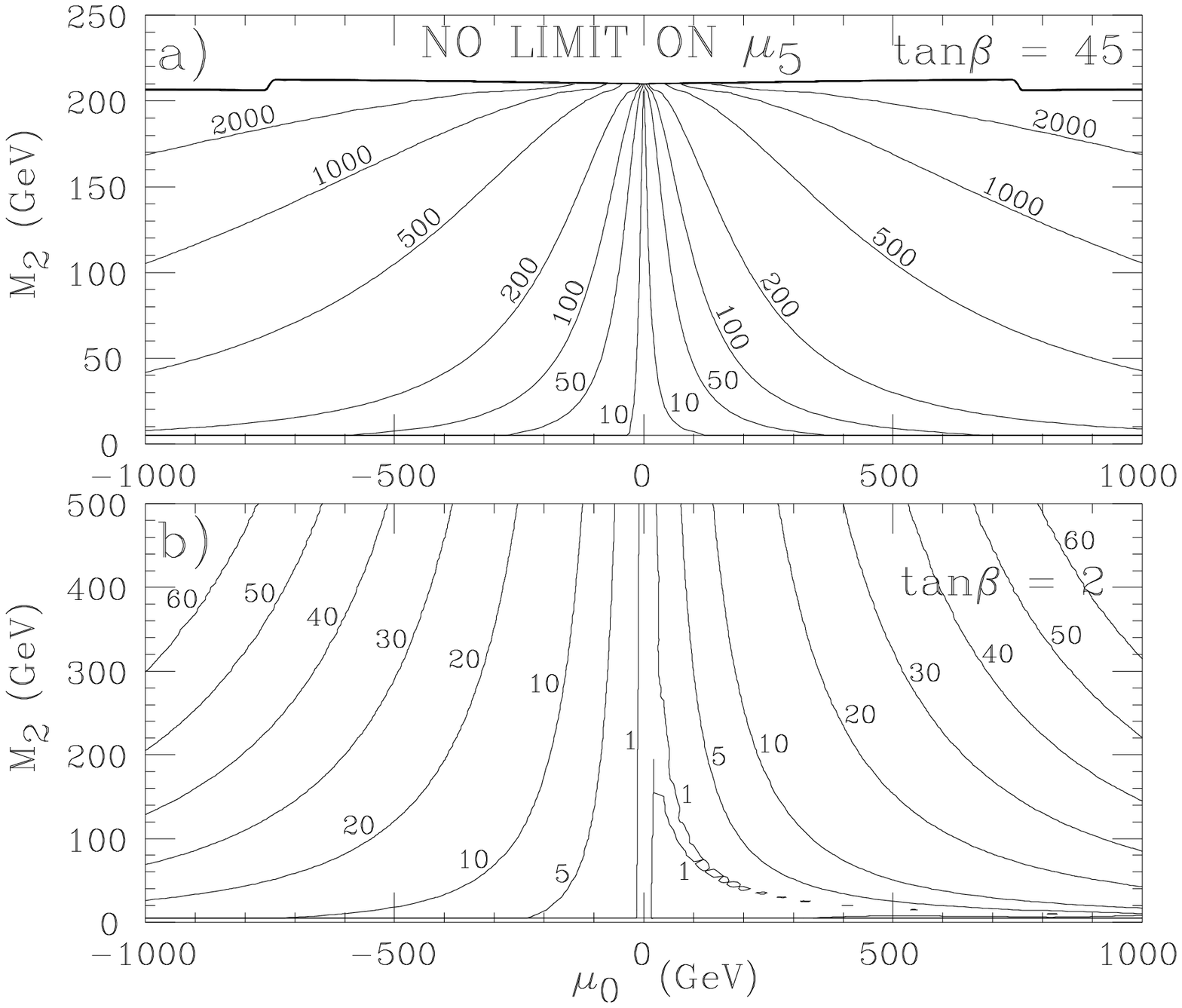}
\caption{
Maximum allowed values of ${\mu}_{\scriptscriptstyle 5}$ (in GeV) consistent 
with $m_{{\nu}_{\tau}} < 24\, \hbox{MeV}$
\newline
(${\mu}_{\scriptscriptstyle 1}:{\mu}_{\scriptscriptstyle 2}:
{\mu}_{\scriptscriptstyle 3} = 0:0:1$).
$M_{\scriptscriptstyle 1} = xM_{\scriptscriptstyle 2}$, with 
$x = \frac{5}{3}{\tan}^2{\theta}_w$ assumed from gaugino unification 
\newline
($M_{\scriptscriptstyle Z} = 91.19\, \hbox{GeV}, \sin\!^2{\!\theta}_w = 0.23$).
The region above or outside of a given contour 
is excluded for ${\mu}_{\scriptscriptstyle 5}$'s above the indicated value.
}

\end{figure}

\eject

\begin{figure}

\vspace{1in}
\includegraphics{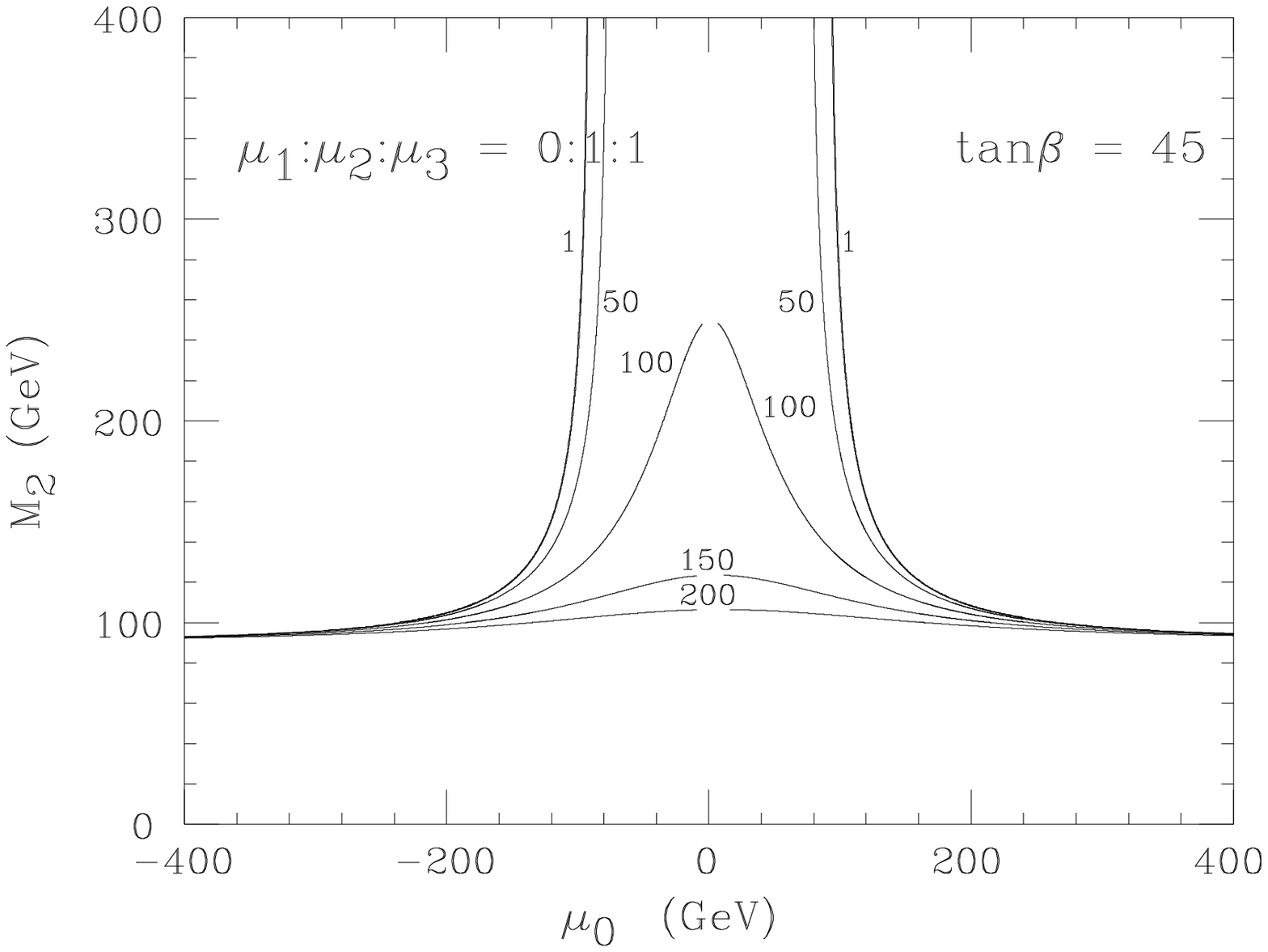}
\caption{
Minimum values of ${\mu}_{\scriptscriptstyle 5}$ (in GeV) required to give
both chargino masses above $90\, \hbox{GeV}$
\newline
(${\mu}_{\scriptscriptstyle 1}:{\mu}_{\scriptscriptstyle 2}
:{\mu}_{\scriptscriptstyle 3} = 0:1:1$).
The area above or outside of a given contour  
has both chargino masses $> \; 90\, \hbox{GeV}$ for 
${\mu}_{\scriptscriptstyle 5}$'s above the indicated value.}

\end{figure}

\eject

\begin{figure}

\vspace{1in}
\includegraphics{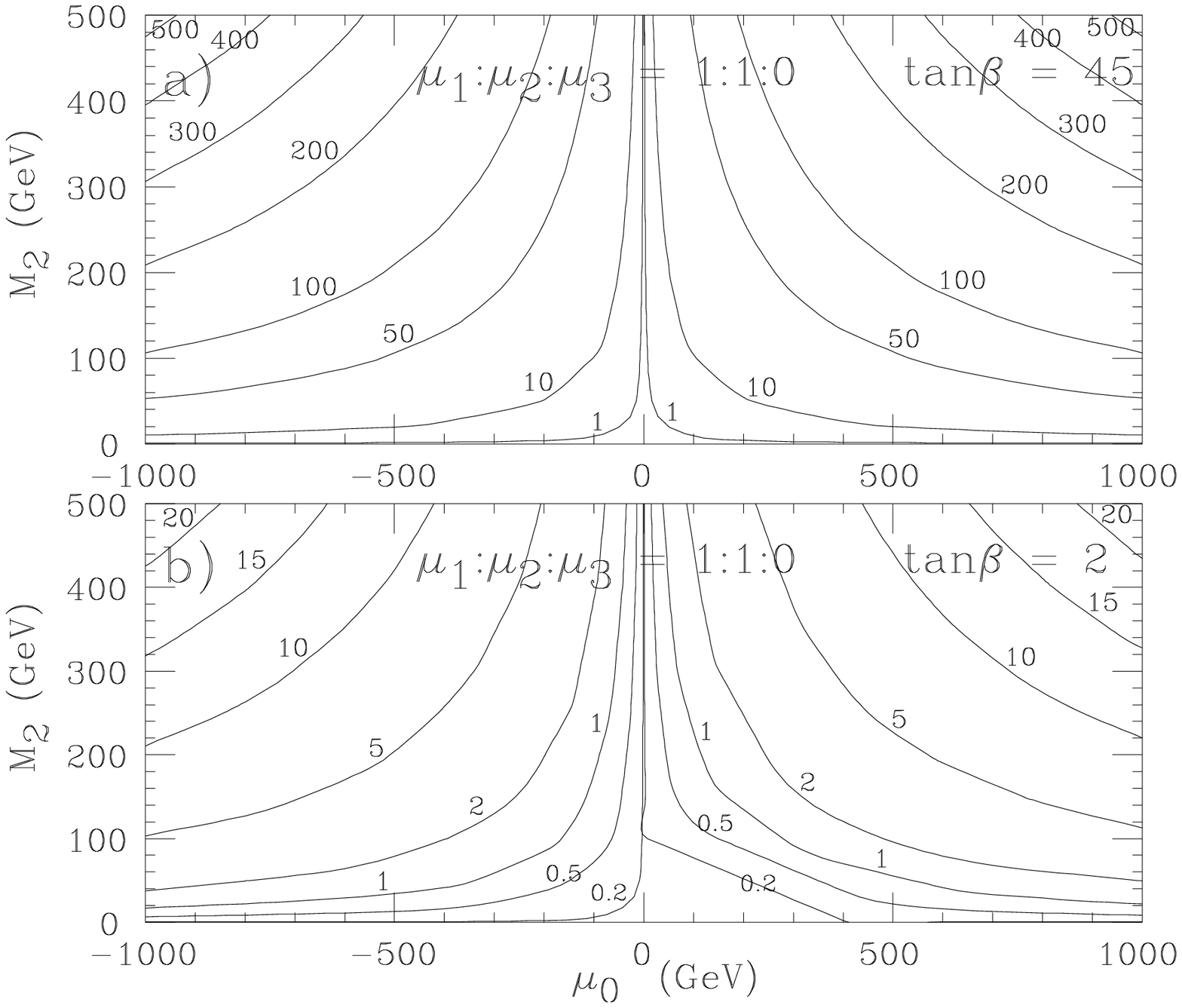}
\caption{
Maximum allowed values of ${\mu}_{\scriptscriptstyle 5}$ (in GeV) consistent 
with B.R.$(\mu^- \longrightarrow e^-e^+e^-) < 1.0^{-12}$ 
\newline
(${\mu}_{\scriptscriptstyle 1}:{\mu}_{\scriptscriptstyle 2}
:{\mu}_{\scriptscriptstyle 3} = 1:1:0$).
The region above or outside of a given contour 
is excluded for ${\mu}_{\scriptscriptstyle 5}$'s above the indicated value.
}

\end{figure}

\end{document}